\documentclass[prb,a4paper]{revtex4}
\usepackage{amsmath} 
\usepackage{amsfonts}
\usepackage{amssymb}
\usepackage{bm} 
\usepackage{graphicx}
\usepackage[applemac]{inputenc}
\usepackage{hyperref}
\usepackage{epstopdf}
\begin{document}
\title{Fermi liquid theory of Fermi-Bose mixtures}
\author{E. V. Thuneberg}
\affiliation{Department of Physics,
P.O.Box 3000, FI-90014 University of Oulu, Finland}
\author{T. H. Virtanen}
\affiliation{Department of Physics,
P.O.Box 3000, FI-90014 University of Oulu, Finland}
\date{\today} 
\begin{abstract}
We write down the basic equations of Fermi-liquid theory for mixtures of fermions and bosons, an example being
$^3$He-$^4$He mixtures at low temperatures. Basically the theory is identical to the one derived by Khalatnikov, but it is derived in a different way, and includes more discussion. A simplifying transformation of the equations is found where the coupling of the normal and superfluid components appears in a simple form. The boundary conditions are discussed.
 \end{abstract}
%
\maketitle
\section{Introduction}
The  Fermi liquid theory formulated by Landau has become a paradigm of what can be the effect of interactions in a Fermi system \cite{Landau57}. Landau formulated the theory originally for liquid $^3$He. Khalatnikov generalized this theory to mixtures of fermions and bosons, and applied it to mixtures of $^3$He and $^4$He [\onlinecite{khal}]. The purpose of this article is to reformulate Khalatnikov's theory. Although the theory is basically the same, our approach is different.  We avoid several complications by concentrating on the laboratory frame, by directly starting with the linearized theory, and by using the osmotic energy. We construct a transformation that eliminates the coupling between the superfluid velocity and the quasiparticle momentum. We discuss boundary conditions.  We also try to interpret the theory by discussing how the momentum of a quasiparticle can be divided into three contributions. The purpose is to formulate general Fermi-Bose liquid equations that are needed to calculate the response on a vibrating wire \cite{VTletter,VTres}.

A general introduction to Fermi liquid theory can be found in Refs.\ \onlinecite{AbrikosovKhalatnikov,Nozieres,LLs2,BP}.
Although our emphasis is on subjects not found in these reviews, we try to be self-contained.
Note that Khalatnikov's theory \cite{khal} as well as the present
theory  make no assumption about the diluteness of the fermion component
in the mixture, which is assumed in several other treatments  \cite{Bardeen67,EE,BP}. Thus the Fermi-liquid theory of a pure fermion system can be obtained as a limiting case of the theory.

We start in Sec.\ \ref{s.bd} with some basic definitions and introduce the noninteracting system.
The basic assumptions of the Fermi-Bose liquid theory are given in Sec.\ \ref{s.ba}. In Sec.\ \ref{s.par} we define the phenomenological parameters that enter the theory. The equations of motion are formulated in Sec.\ \ref{s.em}.  The conserved currents are identified in Sec.\ \ref{s.cur} and some discussion is given. In Sec.\ \ref{s.eli} we make a change to new variables where the coupling between the normal and superfluid components appear only through their densities. The scattering of quasiparticles in the bulk and at surfaces is discussed in Secs.\ \ref{s.col} and \ref{s.bc}. The hydrodynamic limit of the theory is discussed in Sec.\ \ref{s.hd}.

\section{Basic definitions}\label{s.bd}

An important thing to realize is that the {\em mass current density} $\bm J$ is the same as the {\em momentum density} $\bm P$. The equivalence $\bm J\equiv \bm P$ follows because the momentum of a particle is mass times the velocity, $\bm p=m\bm v$, i.e. the mass $m$ is transported at velocity $\bm v$. This relation  is valid in condensed matter under standard (nonrelativistic) conditions since the changes of momentum and mass associated with the interaction field are negligible. This is valid also in crystalline material, where even in the idealized limit of infinitely rigid lattice, one should allow part of the momentum or mass current carried by the lattice. The equivalence $\bm J\equiv \bm P$ is essential in connection of equations (\ref{e.cont3})-(\ref{e.momcons}) below. 

We consider a system of particles having one type of fermions and one type of bosons, with masses $m_{\rm F}$ and $m_{\rm B}$. 
In this section, we describe the system {\em in the absence of interactions}. The free fermions have momenta $\bm p$ and energies $\epsilon_{p}=p^2/2m_{\rm F}$. The state of the fermion system is described by distribution function $n_{\bm p}$ that takes values $0$ and $1$ for each momentum state. 
The number density of fermions $n_{\rm F}$ is given by 
\begin{eqnarray}
n_{\rm F}=\int  n_{\bm p}d\tau,
\label{e.numbdf}\end{eqnarray}
where $d\tau=2d^3 p/(2\pi \hbar)^3$. The fermions have a spin but since we are not considering spin-depedent phenomena, it only appears as a factor of 2 in $d\tau$. 
The ground state consists of a Fermi sphere, which has filled states, $n_{\bm p}=1$,  for momenta  $p<p_{\rm F}$ and empty states, $n_{\bm p}=0$,  for momenta  $p>p_{\rm F}$. The Fermi momentum $p_{\rm F}$ is determined by the number density $n_{\rm F}=p_{\rm F}^3/3\pi^2\hbar^3$. The momentum density of fermions is $\bm J_{\rm F}=\int  \bm p n_{\bm p}d\tau$.
The bosons are assumed to be condensed to a state with velocity $\bm v_{\rm B}$ and all excited states of the boson system are neglected.
The number density of bosons is denoted by $n_{\rm B}$ and the momentum density $\bm J_{\rm B}=m_{\rm B}n_{\rm B} \bm v_{\rm B}$. The total momentum density of both fermions and bosons $\bm J=\bm  J_{\rm F}+\bm J_{\rm B}$ is thus given by
\begin{eqnarray}
\bm J=m_{\rm B}n_{\rm B}\bm v_{\rm B}+\int \bm p\, n_{\bm p}d\tau.
\label{e.jtomacd}\end{eqnarray}

\section{Basic assumptions}\label{s.ba}

We now turn to the discussion of the interacting system. Landau's Fermi-liquid theory applies to low energy excitations of the system.
The basic assumption is  that when interactions are turned on, the low-energy part of the excitation spectrum remains qualitatively the same as it is in the non-interacting system. More precisely, one assumes that the {\em momenta of the excitations remain the same} when the interactions are turned on at constant densities of both fermions and bosons. The energies of the excitations can be shifted, but {\em the equilibrium Fermi surface is assumed to remain unchanged}. Also {\em the quasiparticle  energy is assumed to be linear in $p$} close to the Fermi surface.  Thus the excited states can still be specified by a quasiparticle distribution function $n_{\bm p}$, and the ground state corresponds to the filled Fermi sphere. 

A particular consequence of the assumptions is that Eq.\ (\ref{e.numbdf}) remains valid in the interacting system. For the total momentum of an arbitrary interacting state we write, following Khalatnikov \cite{khal}
\begin{eqnarray}
\bm J=m_{\rm B}n_{\rm B} \bm v_s+\int \bm p \,n_{\bm p}d\tau.
\label{e.jtomacds}\end{eqnarray}
This is the same as Eq.\ (\ref{e.jtomacd}) except that instead of the velocity of bosons it defines {\em the superfluid velocity} $\bm v_s$.

In order to stress the non-triviality of Eq.\ (\ref{e.jtomacds}) we mention that it is {\em incorrect} to deduce that the first term would be the momentum density of bosons and the latter that of fermions. The correct decomposition will be given later in equations  (\ref{e.m3j3}) and  (\ref{e.m4j4}). Note that an interacting system does not have a single boson velocity $\bm v_{\rm B}$, and therefore it cannot appear in Eq.\ (\ref{e.jtomacds}). 

In addition we use in the following the spherical symmetry of the system, the conservation laws of particle number, momentum and energy, and some estimates of orders of magnitude.

\section{Parameterization}\label{s.par}

Above we used variables $n_{\bm p}$, $n_{\rm B}$ and $\bm v_s$ to specify the state of the system. In particular, the energy density  could be written $\tilde E(\{n_{\bm p}\},n_{\rm B},\bm v_s)$. Here the curly brackets indicate that $\tilde E$ is a functional of $n_{\bm p}$, i.e. it depends on $n_{\bm p}$ at all values of $\bm p$. With respect to the variable $n_{\rm B}$, it is more convenient to change to the ``osmotic energy'' defined by 
\begin{eqnarray}
E(\{n_{\bm p}\},\mu_{\rm B},\bm v_s)=\tilde E(\{n_{\bm p}\},n_{\rm B},\bm v_s)-n_{\rm B}\mu_{\rm B}-n_{\rm F}\mu_{\rm F}^{(0)}.
\label{e.eosmotic}\end{eqnarray}
Here the term $-n_{\rm B}\mu_{\rm B}$ effects the standard Legendre transformation from the density $n_{\rm B}$ to the chemical potential
$\mu_{\rm B}=\partial \tilde E(\{n_{\bm p}\},n_{\rm B},\bm v_s)/\partial n_{\rm B}$. The term $-n_{\rm F}\mu_{\rm F}^{(0)}$ looks formally similar for the fermions, but is different because the chemical potential of fermions in a non-equilibrium state is not defined. Instead, we define a reference  state. It has the  equilibrium distribution $n_{\bm p}^{(0)}=\Theta(p_{\rm F}-p)$, where $\Theta(x)$ is the step function. This equilibrium is taken to correspond chemical potentials $\mu_{\rm B}^{(0)}$ and $\mu_{\rm F}^{(0)}$, superfluid velocity $\bm v_s=0$ and temperature $T=0$. Thus the osmotic energy (\ref{e.eosmotic}) still depends on the distribution function $n_{\bm p}$, and the effect of the $-n_{\rm F}\mu_{\rm F}^{(0)}$ term is merely a shift of energy so that quasiparticle energies are counted from $\mu_{\rm F}^{(0)}$.

The quasiparticle
energy  is defined as  
\begin{eqnarray}
\epsilon_{\bm p}=\frac{\delta E(\{n_{\bm p}\},\mu_{\rm B},\bm v_s)}{\delta n_{\bm p}},
\end{eqnarray}
where the functional derivative is interpreted as  $\delta E=\int \epsilon_{\bm p}\delta n_{\bm p}d\tau$.
The linearity of the quasiparticle energy  near the Fermi surface is satisfied by the choice 
\begin{eqnarray}
\epsilon_{\bm p}=v_{\rm F}(p-p_{\rm F})+\delta\epsilon_{\bm p}.
\label{e.ep0pde}\end{eqnarray}
The coefficient of $p-p_{\rm F}$ defines the Fermi velocity $v_{\rm F}$. Writing $v_{\rm F}=p_{\rm F}/m^*$ it defines  the effective mass $m^*$.
The second term
 $\delta\epsilon_{\bm p}$ in (\ref{e.ep0pde}) can be written as
\begin{eqnarray}
\delta\epsilon_{\bm p}=(1+\alpha)\delta\mu_{\rm B}+
D{\bm p}\cdot{\bm v}_s+\int f(\bm p,\bm p_1) (n_{\bm p_1}-n_{ p_1}^{(0)})d\tau_1,
\label{e.deltaex0}\end{eqnarray}
where $\delta\mu_{\rm B}=\mu_{\rm B}-\mu_{\rm B}^{(0)}$ is the deviation of the boson chemical potential from its equilibrium value.
We see that $\delta\epsilon_{\bm p}$ is linear in the deviation from equilibrium, and thus appears less important than the first term in (\ref{e.ep0pde}). It was an important observation of Landau that this term still has to be kept in order to make a consistent theory. Being linear in the deviations of $\mu_{\rm B}$, $\bm v_s$ and $n_{\bm p}$ from equilibrium, we see that  $\delta\epsilon_{\bm p}$ (\ref{e.deltaex0}) has the most general form that is allowed by symmetry. For example the most general function of $\bm p$ that is linear in $\bm v_s$ and does not depend on any other direction, has the form $D(p){\bm p}\cdot{\bm v}_s$ with some function $D(p)$. For the same reason $f(\bm p,\bm p_1)$ cannot depend on the directions of $\hat{\bm p}$ and  $\hat{\bm p}_1$ separately but only on their scalar product $\hat{\bm p}\cdot\hat{\bm p}_1$.

We can further simplify $\delta\epsilon_{\bm p}$ (\ref{e.deltaex0}). Since we are considering only excitations near the Fermi surface, we can approximate $\bm p\approx p_{\rm F}\hat{\bm p}$ in the 
$D{\bm p}\cdot{\bm v}_s$ term.  Also, we estimate that  $f(\bm p,\bm p_1)$ changes essentially only on the momentum scale of $p_{\rm F}$. Since  $n_{\bm p}-n_{\bm p}^{(0)}$ is nonzero only in a thin shell around the Fermi surface, we can neglect the dependence of $f(\bm p,\bm p_1)$ on the magnitudes $p$ and $p_1$. Similarly, we can assume $\alpha$ and $D$ as constants not depending on $p$.
Thus $\delta\epsilon_{\bm p}$ depends only on the momentum direction $\hat{\bm p}$, not on the magnitude $p$.
We can now define an energy integrated distribution function 
\begin{eqnarray}
\phi_{\hat{\bm p}}=\int 
(n_{p\hat{\bm p}}-n^{(0)}_p)v_{\rm F}dp,
\label{e. phidef}\end{eqnarray}
where the argument of $n_{\bm p}$ is $\bm p=p\hat{\bm p}$.
We get that $f_{\bm p,\bm p_1}$ depends only on the scalar product $\hat{\bm p}\cdot\hat{\bm p}_1$.
Thus we can expand $f_{\bm p,\bm p_1}$ using Legendre polynomials $P_l(x)$
\begin{eqnarray}
F(\hat{\bm p}\cdot\hat{\bm p}')\equiv\frac{m^*p_{\rm F}}{\pi^2\hbar^3}f(\hat{\bm p}\cdot\hat{\bm p}')
=\sum_{l=0}^\infty
F_lP_l(\hat{\bm p}\cdot\hat{\bm p}'),
\label{e.fullF}\end{eqnarray}
where $P_0(x)=1$, $P_1(x)=x$, etc and $m^*p_{\rm F}/\pi^2\hbar^3$ is the quasiparticle density of states at the Fermi surface. As a result we can write (\ref{e.deltaex0}) into the form 
\begin{eqnarray}
\delta\epsilon_{\hat{\bm p}}=(1+\alpha)\delta\mu_{\rm B}+
Dp_{\rm F}\hat{\bm p}\cdot{\bm v}_s+
\sum_{l=0}^\infty
F_l\langle P_l(\hat{\bm p}\cdot\hat{\bm p}')\phi_{\hat{\bm p}'}\rangle_{\hat{\bm p}'},
\label{e.deltae}\end{eqnarray}
where $\langle \ldots\rangle_{\hat{\bm
p}}$ denotes the average over the unit sphere of $\hat{\bm p}$. 

We have defined parameters $m^*$, $\alpha$, $D$ and $F_l$'s in equations (\ref{e.ep0pde}) and 
(\ref{e.deltae}). There is one constraint between these required by translational invariance. The relation is conveniently stated by expressing $D$ in terms of other parameters,
\begin{eqnarray}
D=1-\frac{m_{\rm F}}{m^*}\left(1+\frac{1}{3}F_1\right).
\label{e.Dfom}\end{eqnarray}
This relation will be justified below in connection of equation (\ref{e.j3phi}). Note that because of using the osmotic energy (\ref{e.eosmotic}) right from the start, we need to consider only one set of $F_l$ parameters, in contrast to the three sets used in Ref.\ \onlinecite{khal}.

We also need to consider changes in the boson density. We parameterize
\begin{eqnarray}
\delta n_{\rm B}\equiv n_{\rm B}-n_{\rm B}^{(0)}=-(1+\alpha)\delta n_{\rm F}+ \frac{n_{\rm B}}{m_{\rm B} s^2}\delta \mu_{\rm B},
\label{e. dn4dn3dmu4}\end{eqnarray}
where $\delta n_{\rm F}=n_{\rm F}-n_{\rm F}^{(0)}$.
The equality of the coefficient $1+\alpha$ with the one in (\ref{e.deltaex0}) and (\ref{e.deltae})  follows because they are second partial derivatives of $E$ (\ref{e.eosmotic}), 
\begin{eqnarray}
\frac{\partial^2 E}{\partial n_{\rm F}\partial \mu_{\rm B}}
= \frac{\partial  \mu_{\rm F}}{\partial \mu_{\rm B} }=-\frac{\partial  n_{\rm B}}{\partial n_{\rm F} }=1+\alpha.
\end{eqnarray}
In the limit of vanishing concentration of fermions, $\alpha$ reduces to the parameter $\alpha$ used in 
Refs.\ \onlinecite{Bardeen67,BP}. In the same limit, the parameter $s$ in (\ref{e. dn4dn3dmu4}) reduces to the velocity of sound [as will be evident from equations (\ref{e.vseqm}) and (\ref{e.dmuvsd}) below].

\section{Equations of motion}\label{s.em}

In general the distribution function $n_{\bm p}$ depends on location and time, $n_{\bm p}(\bm r,t)$, and similarly for other variables $\mu_{\rm B}(\bm r,t)$ and $\bm v_s(\bm r,t)$. The quasiparticle distribution obeys the kinetic equation 
\begin{eqnarray}
\frac{\partial n_{\bm p} }{\partial  t}
+\bm \nabla n_{\bm p} \cdot\frac{\partial \epsilon_{\bm p} }{\partial  \bm p}-\frac{\partial n_{\bm p} }{\partial  \bm p}\cdot\bm \nabla  \epsilon_{\bm p} =I_{\bm p},
\label{e.kinnp}\end{eqnarray}
where $I_{\bm p}$ is the collision term.
The fermion number and momentum conservation in collisions requires that
\begin{eqnarray}
\int I_{\bm p}d\tau =0,\ \ \ \ \ \int \bm pI_{\bm p}d\tau =0.
\label{e.cosum}\end{eqnarray}
Assuming small deviation from equilibrium, we linearize the kinetic equation (\ref{e.kinnp})  and get
\begin{eqnarray}
\frac{\partial n_{\bm p}}{\partial t}
+v_{\rm F}\hat{\bm p}\cdot\bm\nabla\left(n_{\bm p}-\frac{\partial n_p^{(0)}}{\partial\epsilon_p}\delta\epsilon_{\hat{\bm p}}\right)=I_{\bm p},
\label{e.lb0}\end{eqnarray}
where the derivative ${\partial n_p^{(0)}}/{\partial\epsilon_p}$ should be evaluated at the unperturbed energy $\epsilon_p^{(0)}=v_{\rm F}(p-p_{\rm F})$.

The superfluid velocity $\bm v_s$ appearing in (\ref{e.jtomacds}) is assumed to be curl free \cite{Kbook}
\begin{eqnarray}
\bm\nabla\times{\bm v}_s=0.
\label{e.curlvs0}\end{eqnarray}
It is assumed to obey the ideal-fluid equation of motion,
\begin{eqnarray}
\frac{\partial {\bm v}_s}{\partial t}+\frac1{m_{\rm B}}\bm\nabla\mu_{\rm B}=0.
\label{e.vseqm}\end{eqnarray}

\section{Currents}\label{s.cur}

Based on the equations above one should be able to derive conservation laws. In particular, one should be able to derive separate  continuity equations for the two components
\begin{eqnarray}
m_{\rm F}\frac{\partial n_{\rm F}}{\partial t}+\bm\nabla\cdot{\bm J}_{\rm F}=0,\label{e.cont3}\\
m_{\rm B}\frac{\partial n_{\rm B}}{\partial t}+\bm\nabla\cdot{\bm J}_{\rm B}=0,\label{e.cont4}
\end{eqnarray}
and the momentum conservation
\begin{eqnarray}
\frac{\partial {\bm J}}{\partial t}+\bm\nabla\cdot\tensor\Pi=0.
\label{e.momcons}\end{eqnarray}
where $\bm J=\bm J_{\rm F}+\bm J_{\rm B}$. In the following we derive linearized expressions for the fermion mass current $\bm J_{\rm F}$, for the boson mass current $\bm J_{\rm B}$ and for the momentum flux tensor $\tensor\Pi$.

We integrate the kinetic equation (\ref{e.lb0}) over all momenta. Using (\ref{e.numbdf})  and  (\ref{e.deltae}) and comparing to (\ref{e.cont3}) we
get
\begin{eqnarray}
{\bm J}_{\rm F}=Dm_{\rm F}n_{\rm F}{\bm v}_s+\frac{m_{\rm F}}{m^*}(1+\frac{1}{3}F_1)\int{\bm p}\,(n_{\bm p}-n_{ p}^{(0)})d\tau.
\label{e.m3j3}\end{eqnarray}
From equations (\ref{e.jtomacds}) and (\ref{e.m3j3}) we can eliminate the integral term and get
\begin{eqnarray}
\bm J=\left(m_{\rm B}n_{\rm B}-\frac{Dm^*n_{\rm F}}{1+\frac13F_1}\right)\bm v_s+
\frac{m^*}{m_{\rm F}(1+\frac13F_1)}\bm J_{\rm F}.
\label{e.j3phi}\end{eqnarray}
This can be used to justify relation (\ref{e.Dfom}). Consider equilibrium but the whole system moving at constant velocity $\bm v$.
Because it has to be that $\bm J=(m_{\rm F}n_{\rm F}+m_{\rm B}n_{\rm B})\bm v$, $\bm v_s=\bm v$ and $\bm J_{\rm F}=m_{\rm F}n_{\rm F}\bm v$, we get  (\ref{e.Dfom}).

Using (\ref{e.jtomacds}) and (\ref{e.m3j3}) we calculate $\bm J_{\rm B}=\bm J-\bm J_{\rm F}$ and get
\begin{eqnarray}\label{current4}
\bm J_{\rm B}=(m_{\rm B}n_{\rm B}-Dm_{\rm F}n_{\rm F}){\bm v}_s+D\int{\bm p}\,(n_{\bm p}-n_{p}^{(0)})d\tau.
\label{e.m4j4}\end{eqnarray}
Let us consider Eqs.\ (\ref{e.m3j3}) and (\ref{e.m4j4}) in the frame where $\bm v_s=0$. From Eq.\ (\ref{e.m4j4}) we see that a single quasiparticle of momentum $\bm p$ has fraction $D$ of the momentum carried by bosons. As the principal fermion forming the quasiparticle travels with velocity $v=p/m^*$, it carries the fraction $m_{\rm F}/m^*$ of the quasiparticle momentum. The rest fraction of the momentum $m_{\rm F}F_1/3m^*$ is carried by other fermions. The total fermion contribution is visible in the second term of $\bm J_{\rm F}$ (\ref{e.m3j3}). 

In order to clarify the nature of the quasiparticle, we consider a simple model. Assume that the principal fermion forms a sphere of radius $a$ and the bosons and other fermions are modeled as an incompressible ideal fluid of density $\rho$. In this model one can calculate $m^*=m_{\rm F}+2\pi a^3\rho/3$, i.e.\ the effective mass  is the fermion mass plus the fluid mass corresponding to half of the sphere. This result is obtained by calculating the momentum $\bm p=m^*\bm v$ from Newton's second law when a force is applied to accelerate the sphere, or, equivalently, by calculating the kinetic energy $E_k$ and writing $E_k=\frac12m^*v^2$.   This model gives a simple picture how the effective mass is increased due to the medium dragged along by the principal fermion. 
This is sometimes called ``backflow'' but this may be misleading since the momentum  is in the ``forward'' direction (as $m^*> m_{\rm F}$). 

We note that when considering a single excited quasiparticle in a finite container, one must also allow a compensating flow. 
Using the simple quasiparticle model as discussed above, the total momentum of an incompressible fluid in finite stationary container must vanish. Thus the momentum $\bm p=m^*\bm v$ of a quasiparticle must be compensated by an opposite flow of the medium. This compensating momentum arises from force applied by the walls of the container, and thus is separate from the momentum of the quasiparticle.

For further insight, consider the case that the fermions are in equilibrium but moving with velocity $\bm v$, i.e. $\bm J_{\rm F}=m_{\rm F}n_{\rm F}\bm v$. We assume  $\bm v_s=0$.  The mass density dragged in such a
normal current  is called {\em normal fluid density} $\rho_n$, i.e. $\bm J=\rho_n\bm v$. The complementary fraction of the total density $\rho=m_{\rm F}n_{\rm F}+m_{\rm B}n_{\rm B}$  that remains at rest, is called  {\em superfluid density} $\rho_s=\rho-\rho_n$. From (\ref{e.j3phi}) we get
\begin{eqnarray}
\rho_s=m_{\rm B}n_{\rm B}-\frac{Dm^*n_{\rm F}}{1+F_1/3},\ \ \
\rho_n=\frac{m^*n_{\rm F}}{1+F_1/3}.
\label{e.rhosnh}\end{eqnarray}
These expressions have a straightforward interpretation in terms of the momentum division introduced  above. Namely $\bm J_{\rm F}=m_{\rm F}(1+\frac13F_1)\bm J/m^*$ and 
$\bm J_{\rm B}=D\bm J$. Solving $\bm J$ from the former and using $\bm J_{\rm F}=m_{\rm F}n_{\rm F}\bm v$ one gets the normal fluid density  (\ref{e.rhosnh}). One also sees that the mass density $Dm^*n_{\rm F}/(1+F_1/3)$ subtracted from the boson density in $\rho_s$ (\ref{e.rhosnh}) just corresponds to the boson fraction that is bound to the quasiparticles. 

Next we check the momentum conservation (\ref{e.momcons}). 
We take the time derivative of ${\bm J}$ (\ref{e.jtomacds}), and use the kinetic equation of quasiparticles (\ref{e.lb0}) and the equation of motion of the Bose component (\ref{e.vseqm}).  Using (\ref{e.cosum}) one finds the momentum conservation with the momentum flux tensor 
\begin{eqnarray}
\tensor\Pi=P^{(0)} \tensor 1+n_{\rm B}\delta\mu_{\rm B}\tensor 1+\frac1{m^*}
\int{\bm p}\,{\bm p}\,\left(n_{\bm p}-n_{p}^{(0)}-\frac{\partial n_p^{(0)}}{\partial\epsilon_p}\delta\epsilon_{\hat{\bm p}}\right)d\tau.
\label{e.mftenw}\end{eqnarray}
Here the constant term arising from  the equilibrium pressure $P^{(0)}$ is added.

\section{Elimination of  superfluid-velocity coupling}\label{s.eli}

Our purpose in this section is to rewrite the theory so that the coupling between the superfluid and normal fluid components appears in a minimal way.
For that purpose we define new quantities $\delta\bar n_{\bm p}$ and $\delta\bar \epsilon_{\bm p}$, and
write the kinetic equation in the form
\begin{eqnarray}
\frac{\partial }{\partial t}\left(\delta \bar n_{\bm p}+\frac{\partial n^{(0)}}{\partial \epsilon_{p}}\delta\bar\epsilon_{\hat{\bm p}}\right)
+v_{\rm F}\hat{\bm p}\cdot\bm\nabla\delta \bar n_{\bm p}=I_{\bm p},
\label{e.lbtf}\end{eqnarray}
so that $\delta\bar\epsilon_{\bm p}$ will only depend on $\delta\mu_{\rm B}$ and on $\delta \bar n_{\bm p}$ but not on $\bm v_s$. In order to achieve this, we scalar multiply the superfluid equation (\ref{e.vseqm}) by $Av_{\rm F}\hat{\bm p}\,{\partial n^{(0)}}/{\partial \epsilon_{p}}$ and add it to the kinetic equation (\ref{e.lb0}). $A$ is a constant that is fixed below. Comparing this to (\ref{e.lbtf}) we find 
\begin{eqnarray}
\delta \bar n_{\bm p}=n_{\bm p}-n_{p}^{(0)}-\frac{\partial n^{(0)}}{\partial \epsilon_{p}}\left(\delta\epsilon_{\hat{\bm p}}+\frac{A}{m_{\rm B}}\delta\mu_{\rm B}\right),\label{e.dbarnp}\\
\delta\bar\epsilon_{\hat{\bm p}}=\delta\epsilon_{\hat{\bm p}}+\frac{A}{m_{\rm B}}\delta\mu_{\rm B}-Av_{\rm F}\hat{\bm p}\cdot\bm v_s.
\end{eqnarray}
Similar to $\phi_{\hat{\bm p}}$ (\ref{e. phidef}), it is convenient to define
\begin{eqnarray}
\psi_{\hat{\bm p}}=\int 
\delta\bar n_{p\hat{\bm p}}v_{\rm F}dp.
\label{e. psidef}\end{eqnarray}
The next task is to express $\delta\bar\epsilon_{\hat{\bm p}}$ in terms of $\psi_{\hat{\bm p}}$. Doing this one notices that the dependence on $\bm v_s$ drops out by choosing $A=Dm^*/(1+F_1/3)$ and we get
\begin{eqnarray}
\delta\bar\epsilon_{\hat{\bm p}}
=\frac{K}{1+ F_0}\delta\mu_{\rm B}
+\sum_{l=0}^\infty
\frac{ F_l}{1+\frac{1}{2l+1} F_l}
\langle P_l(\hat{\bm p}\cdot\hat{\bm p}')\psi_{\hat{\bm p}'}\rangle_{\hat{\bm p}'},
\label{e.deltae2}\end{eqnarray}
where we have defined
\begin{eqnarray}
K=\frac{m^*D}{m_{\rm B}(1+\frac13F_1)}+1+\alpha
=\frac{m^*}{m_{\rm B}(1+\frac13F_1)}-\frac{m_{\rm F}}{m_{\rm B}}+1+\alpha.
\end{eqnarray}

It is useful to express the densities and currents in terms of $\delta\bar n_{\bm p}$ (\ref{e.dbarnp}) or $\psi_{\hat{\bm p}}$ (\ref{e. psidef}).  Here we give  the fermion density (\ref{e.numbdf}) 
\begin{eqnarray}
\delta n_{\rm F}\equiv  n_{\rm F}-n_{{\rm F}}^{(0)}=\frac{1}{1+ F_0}\left(\int\delta\bar n_{\bm p}d\tau-
\frac{m^*p_{\rm F} K}{\pi^2\hbar^3}\delta\mu_{\rm B}\right)=\frac{m^*p_{\rm F}}{\pi^2\hbar^3(1+ F_0)}
\left(\langle\psi_{\hat{\bm p}}\rangle_{\hat{\bm p}}-K\delta\mu_{\rm B}\right),
\label{e.n3dens}\end{eqnarray}
the fermion momentum (\ref{e.m3j3}) 
\begin{eqnarray}
{\bm J}_{\rm F}=\frac{m_{\rm F}}{m^*}
\int{\bm p}\,\delta\bar n_{\bm p}d\tau
=
\frac{m_{\rm F}p_{\rm F}^2}{\pi^2\hbar^3}
\langle\hat{\bm p}\psi_{\hat{\bm p}}\rangle_{\hat{\bm
p}},
\label{e.currt}\end{eqnarray}
 and the momentum flux tensor (\ref{e.mftenw}) 
\begin{eqnarray}
\tensor\Pi=P^{(0)} \tensor 1+\frac{\rho_s}{m_{\rm B}}\delta\mu_{\rm B}\tensor 1+\frac1{m^*}
\int{\bm p}\,{\bm p}\,\delta\bar n_{\bm p}d\tau=P^{(0)} \tensor 1+\frac{\rho_s}{m_{\rm B}}\delta\mu_{\rm B}\tensor 1+3n_{\rm F}\langle\hat{\bm p}\hat{\bm p}\psi_{\hat{\bm p}}\rangle_{\hat{\bm
p}}.
\label{e.mftenw2}\end{eqnarray}
In the momentum flux (\ref{e.mftenw2})  the  second and third terms arise from the superfluid and the normal components, respectively.
We also rewrite the boson conservation (\ref{e.cont4}) using (\ref{e. dn4dn3dmu4}) as
\begin{eqnarray}
\frac{n_{{\rm B}}}{s^2}\frac{\partial\delta\mu_{\rm B}}{\partial t}+\rho_s\bm\nabla\cdot\bm v_s-m_{\rm B}K\frac{\partial n_{\rm F}}{\partial t}=0.
\label{e.dmuvsd}\end{eqnarray}
The equations  (\ref{e.lbtf}), (\ref{e.dmuvsd}), and (\ref{e.vseqm}) constitute equations of motion in variables $\delta\bar n_{\bm p}$,  $\delta\mu_{\rm B}$, and $\bm v_s$.

We see that in the system of equations (\ref{e.vseqm}), (\ref{e.lbtf}), (\ref{e.deltae2}), and (\ref{e.dmuvsd}), the coupling of the superfluid and normal components takes place through their densities via the terms proportional to 
$K$. Let us estimate this effect considering a problem of length scale $a$, frequency $\omega$, and relatively long mean free path $\ell \gtrsim a$. We first consider fermion flow with velocity scale $u$, for which we estimate $\delta n_{\rm F}\sim n_{\rm F}u/v_{\rm F}$. Assuming $\omega a/s\ll 1$, the first term in (\ref{e.dmuvsd}) can be neglected, and it is important to balance the last two terms.  We get that a fermion flow with velocity $u$ generates a superfluid velocity of order $K(n_{\rm F}/n_{\rm B})(\omega a/v_{\rm F})u$. The opposite effect of $v_s\sim u$ on the quasiparticles can be estimated from the fermion conservation (\ref{e.cont3}) and applying (\ref{e.n3dens}) and (\ref{e.currt}). We get that for $\omega a/v_{\rm F}\lesssim1$  the induced fermion velocity is of order  $K(m_{\rm B}/m_{\rm F})(\omega a/v_{\rm F})^2u$. For small frequencies  $\omega a/v_{\rm F}\ll1$ the velocities induced in the other component are small, and the superfluid and quasiparticle components can be treated independently of each other.

\section{Collision term}\label{s.col}

For detailed discussion of the collision term $I_{\bm p}$ we refer to Sykes and Brooker  \cite{Sykes70}. In the collision of two quasiparticles, their total energy is conserved. It follows that the collision integral can be written as a function $\delta\bar n_{\bm p}$ (\ref{e.dbarnp}). The collision term  can in a few cases be analyzed exactly \cite{Sykes70}. A simpler approach is to use relaxation-time approximation. In its simplest version one assumes \cite{BK}
\begin{eqnarray}
I_{\bm p}=-\frac1\tau\left[n_{\bm p}-n^{(0)}(\epsilon^{(0)}_{p}+\delta\epsilon_{\bm p}-c-\bm b\cdot\bm p)\right],
\end{eqnarray}
where $\tau$ is the relaxation time and coefficients $c$ and $\bm b$ are chosen so that  conditions (\ref{e.cosum}) are satisfied. The relaxation time can also be specified  by the mean free path $\ell=v_{\rm F}\tau$. The relaxation-time approximation leads to essential simplification because instead of using $\delta n_{\bm p}$ or $\delta\bar n_{\bm p}$, one can construct equations for $\psi_{\hat{\bm p}}$ (\ref{e. psidef}), which does not depend on the magnitude of the momentum. In the relaxation-time approximation the kinetic equation (\ref{e.lbtf}) takes the form 
\begin{eqnarray}
\frac{\partial}{\partial t}(\psi_{\hat{\bm p}}-\delta\bar\epsilon_{\hat{\bm p}})
+v_{\rm F}\hat{\bm p}\cdot\bm\nabla\psi_{\hat{\bm p}}=
-\frac{1}{\tau}\left(\psi_{\hat{\bm p}}-\langle\psi_{\hat{\bm p}'}\rangle_{\hat{\bm p}'}
-3\langle\hat{\bm p}\cdot\hat{\bm p}'\psi_{\hat{\bm p}'}\rangle_{\hat{\bm p}'}
\right).
\label{e.lbrel2}\end{eqnarray}
It is also possible to introduce different relaxation times $\tau_l$ (with $l=2,3,\ldots$) corresponding to spherical harmonic decomposition of $\psi_{\hat{\bm p}}$ (\ref{e.shdpsi}).

\section{Boundary conditions}\label{s.bc}

The boundary conditions at walls depend on the scattering properties of the wall. Two common models for surface scattering are specular and diffuse scattering. We generalize the analysis of Ref.\ \onlinecite{BK} to Fermi-Bose liquids and include general interactions (\ref{e.fullF}).
We first consider the boundary conditions in the rest frame of a wall, and after that generalize to the case of a moving wall. 

We assume a wall with normal $\hat{\bm n}$ that is inpenetrable to both bosons and fermions. From equation (\ref{e.j3phi}) we see that the vanishing of the particle fluxes $\bm J$ and $\bm J_{\rm F}$ normal to the wall imply vanishing of the normal component of the superfluid velocity,
\begin{eqnarray}
\hat{\bm n}\cdot\bm v_s=0.
\label{e.bcvs0}\end{eqnarray}

Specular scattering means that the quasiparticle momentum  changes from $\bm p$ to 
$\bm p_R=\bm p-2\hat{\bm n}(\hat{\bm n}\cdot\bm p)$, and thus
\begin{eqnarray}
n_{\bm p}=n_{\bm p-2\hat{\bm n}(\hat{\bm n}\cdot\bm p)}.
\label{spec bcn}\end{eqnarray}
Before accepting this one should check that it conserves the quasiparticle energy (\ref{e.ep0pde}), i.e.\ we calculate
\begin{eqnarray}
\delta\epsilon_{\hat{\bm p}}-\delta\epsilon_{\hat{\bm p}_R}=
Dp_{\rm F}(\hat{\bm p}-\hat{\bm p}_R)\cdot{\bm v}_s+
\langle F(\hat{\bm p}\cdot \hat{\bm p}')\phi_{\hat{\bm p}'}-F(\hat{\bm p}_R\cdot \hat{\bm p}')\phi_{\hat{\bm p}'}\rangle_{\hat{\bm p}'}\nonumber\\
=2Dp_{\rm F}(\hat{\bm n}\cdot \hat{\bm p})(\hat{\bm n}\cdot{\bm v}_s)+
\langle F(\hat{\bm p}\cdot \hat{\bm p}')\phi_{\hat{\bm p}'}-F(\hat{\bm p}_R\cdot \hat{\bm p}_R')\phi_{\hat{\bm p}'_R}\rangle_{\hat{\bm p}'}\nonumber\\
=2Dp_{\rm F}(\hat{\bm n}\cdot \hat{\bm p})(\hat{\bm n}\cdot{\bm v}_s)+
\langle F(\hat{\bm p}\cdot \hat{\bm p}')[\phi_{\hat{\bm p}'}-\phi_{\hat{\bm p}'_R}]\rangle_{\hat{\bm p}'}=0
\end{eqnarray}
where we have used (\ref{e.deltae}), $\hat{\bm p}_R\cdot \hat{\bm p}_R'=\hat{\bm p}\cdot \hat{\bm p}'$, (\ref{e.bcvs0}) and $\phi_{\hat{\bm p}'}-\phi_{\hat{\bm p}'_R}=0$, which follows from (\ref{spec bcn}).

Diffuse scattering means that the quasiparticles reflected from the wall are in equilibrium evaluated at the exact quasiparticle energy (\ref{e.ep0pde}). Thus the distribution of the reflected quasiparticles is
\begin{eqnarray}
n_{\bm p}
=n^{(0)}(\epsilon_{p}^{(0)}+\delta\epsilon_{\bm p}-c)
=n^{(0)}_p+\frac{dn^{(0)}}{d\epsilon_p}(\delta\epsilon_{\bm p}-c)\ \ \ {\rm for}\  \hat{\bm n}\cdot\bm p>0,
\label{e.dbcn0}\end{eqnarray}
where the constant $c$ is determined by the condition that the particle flux (\ref{e.currt}) to the wall vanishes, $\hat{\bm n}\cdot\bm J_{\rm F}=0$. More conveniently, the boundary is expressed as 
\begin{eqnarray}
\delta\bar n_{\bm p}=-\frac{dn^{(0)}}{d\epsilon_p}\bar c\ \ \ {\rm for}\  \hat{\bm n}\cdot\bm p>0
\label{e.dbcn1}\end{eqnarray}
with a new constant $\bar c$. The constant can be determined by evaluating the condition $\hat{\bm n}\cdot\bm J_{\rm F}=0$ for reflected quasiparticles, which  allows to write the diffuse boundary condition as
\begin{eqnarray}
\delta\bar n_{\bm p}=\frac{4\pi^2\hbar^3}{m^*p_{\rm F}}\frac{dn^{(0)}}{d\epsilon_p}
\int_{\hat{\bm n}\cdot\hat{\bm p}'<0} \hat{\bm n}\cdot\hat{\bm p}'\,\delta\bar n_{\bm p'}\, d\tau'\ \ \ {\rm for}\  \hat{\bm n}\cdot\bm p>0.
\label{e.dbcn2}\end{eqnarray}

The boundary conditions in the case of a moving wall can be found by changing the reference frame. In the laboratory frame the quasiparticle momentum $\bm p=\bm p'+m_{\rm F}\bm u$, where $\bm p'$ is the momentum in the rest frame of the wall, and $\bm u$ is the velocity of the wall. (This relation follows because the momenta are assumed to remain unchanged when the interactions are switched on.) 
The quasiparticle distributions in the two frames are the same, $n_{\bm p}=n'_{\bm p'}$. A change arises in $\delta n_{\bm p}$,
\begin{eqnarray}
\delta n_{\bm p}=n_{\bm p}-n_p^{(0)}=n'_{\bm p'}-n_{\bm p'+m_{\rm F}\bm u}^{(0)}=\delta n'_{\bm p'}-\frac{dn^{(0)}}{d\epsilon_p}m_{\rm F}v_{\rm F}\hat{\bm p}\cdot\bm u
\end{eqnarray}
Based on this we find $\phi_{\hat{\bm p}}=\phi'_{\hat{\bm p}}+m_{\rm F}v_{\rm F}\hat{\bm p}\cdot\bm u$, and using $ \bm v_s= \bm v_s'+\bm u$ we get $\delta \epsilon_{\hat{\bm p}}
=\delta \epsilon'_{\hat{\bm p}}+(m^*-m_{\rm F})v_{\rm F}\hat{\bm p}\cdot\bm u$. The transformation for  $\delta\bar n_{\bm p}$ is then 
\begin{eqnarray}
\delta\bar n_{\bm p}=\delta\bar n'_{\bm p'}-\frac{dn^{(0)}}{d\epsilon_p}p_{\rm F}\hat{\bm p}\cdot\bm u
\label{e.tfrnb}
\end{eqnarray}
which implies $\psi_{\hat{\bm p}}=\psi'_{\hat{\bm p}}+p_{\rm F}\hat{\bm p}\cdot\bm u$.

Applying the transformation rules to (\ref{e.bcvs0}), (\ref{spec bcn}) and (\ref{e.dbcn2}) gives the superfluid condition 
\begin{eqnarray}
\hat{\bm n}\cdot{\bm v}_s=\hat{\bm n}\cdot{\bm u},
\label{e.bcvs}\end{eqnarray}
the specular boundary condition
\begin{eqnarray}
\delta\bar n_{\bm p}=\delta\bar n_{\bm p-2\hat{\bm n}(\hat{\bm n}\cdot\bm p)}+\frac{dn^{(0)}}{d\epsilon_p}2p_{\rm F}\hat{\bm n}\cdot\hat{\bm p}\,\hat{\bm n}\cdot\bm u,
\label{e.spbcu}\end{eqnarray}
and the diffusive boundary condition \begin{eqnarray}
\delta\bar n_{\bm p}=4\frac{dn^{(0)}}{d\epsilon_p}\frac{d\epsilon_p}{d\tau}\int_{\hat{\bm n}\cdot\hat{\bm p}'<0} \hat{\bm n}\cdot\hat{\bm p}'\,\delta\bar n_{\bm p'}\, d\tau'
-\frac{dn^{(0)}}{d\epsilon_p}p_{\rm F}(\hat{\bm p}+\frac{2}{3}\hat{\bm n})\cdot{\bm u}
\ \ \ {\rm for}\  \hat{\bm n}\cdot\bm p>0.
\label{e.dbcn3}\end{eqnarray}
In terms of  $\psi_{\hat{\bm p}}$ the specular and diffuse conditions can be written
\begin{eqnarray}\label{spec bc2}
\psi_{\hat{\bm p}}=\psi_{\hat{\bm p}-2\hat{\bm n}(\hat{\bm
n}\cdot\hat{\bm p})}+2p_{\rm F}(\hat{\bm n}\cdot\hat{\bm
p})(\hat{\bm n}\cdot{\bm u}).
\end{eqnarray}
\begin{eqnarray}\label{diff bc}
\psi_{\hat{\bm p}_{\rm out}}=
-2\langle\hat{\bm n}\cdot\hat{\bm p}_{\rm in}
\psi_{\hat{\bm p}_{\rm in}}\rangle_{\hat{\bm
p}_{\rm in}}
+p_{\rm F}(\hat{\bm p}_{\rm
out}+\frac{2}{3}\hat{\bm n})\cdot{\bm u}\end{eqnarray}
with an average over half of the unit sphere ($\hat{\bm n}\cdot\hat{\bm
p}_{\rm in}<0$).

A consistency check for the boundary conditions is that the
fermion particle current (\ref{e.currt})
behaves as expected
\begin{eqnarray}
\hat{\bm n}\cdot{\bm J}_{\rm F}=m_{\rm F}n_{\rm F}\hat{\bm n}\cdot{\bm u}.
\label{e.genbc}\end{eqnarray}
From the specular boundary condition (\ref{spec bc2}) and from (\ref{e.mftenw2}) we deduce the vanishing of the transverse momentum flux,
\begin{eqnarray}
\hat{\bm n}\cdot\tensor\Pi\times\hat{\bm n}=0.
\label{e.bcspec}\end{eqnarray}

\section{Hydrodynamic limit}\label{s.hd}

The kinetic theory reduces to hydrodynamic theory in the limit of small mean free path of quasiparticles.
We sketch the derivation here starting from the relaxation-time approximation.
The quasiparticle distribution function $\psi_{\hat{\bm p}}$ can be expanded in spherical harmonics
\begin{eqnarray}
\psi_{\hat{\bm p}}=\sum_{l=0}^{\infty}\sum_{m=-l}^{l}c_l^mY_l^m(\hat{\bm p}).
\label{e.shdpsi}\end{eqnarray}
The lowest order  coefficients $c_0^0$ and $c_1^m$ can be related to fermion number (\ref{e.n3dens}) and current  (\ref{e.currt}). Due to rapid scattering, all higher order coefficients are assumed to be small. We define {\em normal fluid velocity} $\bm v_n$ by writing the current $\bm J_{\rm F}=n_{\rm F} \bm v_n$.
We multiply the kinetic equation (\ref{e.lbrel2}) by $\hat{\bm p}\hat{\bm p}-\tensor 1/3$ and average over $\hat{\bm p}$. We get
\begin{eqnarray}
\frac{\partial}{\partial t}\langle(\hat{\bm p}\hat{\bm p}-\tensor 1/3)[\psi_{\hat{\bm p}}-\delta\bar\epsilon_{\hat{\bm p}}]\rangle
+v_{\rm F}\langle(\hat{\bm p}\hat{\bm p}-\tensor 1/3)\hat{\bm p}\cdot\bm\nabla\psi_{\hat{\bm p}}\rangle=
-\frac{1}{\tau}\langle(\hat{\bm p}\hat{\bm p}-\tensor 1/3)\psi_{\hat{\bm p}}\rangle.
\label{e.lbrel3}\end{eqnarray}
In the limit of small gradients and time derivatives, the time-derivative term and the coupling to third order term of (\ref{e.shdpsi}) in the gradient term of (\ref{e.lbrel3}) can be neglected. This allows to write the momentum flux tensor (\ref{e.mftenw2}) into the form
\begin{eqnarray}
\Pi_{ij}=P_0\delta_{ij}
+\left(\frac{\rho_s}{m_{\rm B}}\delta\mu_{\rm B}+\delta P^*\right)\delta_{ij}
-\eta
\left(\frac{\partial v_{ni}}{\partial x_j}+\frac{\partial v_{nj}}{\partial x_i}-\frac{2}{3}\bm\nabla\cdot\bm v_n\delta_{ij}\right),
\label{e.moflhy}\end{eqnarray}
where effective pressure of the normal component
\begin{eqnarray}
\delta P^*=Kn_{{\rm F}}\delta\mu_{\rm B} +\frac13m^*v_{\rm F}^2(1+ F_0)\delta n_{\rm F}
\end{eqnarray}
and the coefficient of  viscosity
\begin{eqnarray}
\eta=\frac{1}{5}n_{\rm F}v_{\rm F}p_{\rm F}\tau.
\end{eqnarray}

The momentum conservation
(\ref{e.momcons}) combined with the flux (\ref{e.moflhy}) gives an equation of motion. Combined with the superfluid equation of motion (\ref{e.vseqm}), the fermion conservation law (\ref{e.cont3}) and the boson conservation law (\ref{e.dmuvsd}) gives a complete set of linearized hydrodynamic equations.
Using variables $(\delta n_{\rm F},\delta\mu_{\rm B},\bm v_n,\bm v_s)$  the set can be written as
\begin{eqnarray}
\frac{\partial \delta n_{\rm F}}{\partial t}+n_{\rm F}\bm\nabla\cdot\bm v_n&=&0,\label{e.hydrogroupb1}\\
\frac{n_{\rm B}}{s^2}\frac{\partial \delta\mu_{\rm B}}{\partial t}+\rho_s\bm\nabla\cdot\bm v_s+m_{\rm B}Kn_{\rm F}\bm\nabla\cdot\bm v_n&=&0,\label{e.hydrogroupb2}\\
\rho_n\frac{\partial \bm v_n}{\partial t}+\bm\nabla \delta P^*&=&\eta\nabla^2\bm v_n
+\textstyle{\frac{1}{3}}\eta\bm\nabla\bm\nabla\cdot\bm v_n,
\label{e.hydrogroupb3}\\
m_{\rm B}\frac{\partial \bm v_s}{\partial t}+\bm\nabla\delta\mu_{\rm B}&=&0.\label{e.hydrogroupb4}
\end{eqnarray}
Here (\ref{e.hydrogroupb3}) is the Navier-Stokes equation describing the viscous normal component.

One application of the equations (\ref{e.hydrogroupb1})-(\ref{e.hydrogroupb4}) is to determine the velocities of first and second sound modes. The results are identical to those found by Khalatnikov \cite{khal}.

\section{Discussion}

For {\em stationary} phenomena, the Fermi-Bose liquid theory does not differ from the noninteracting system, when written in terms of proper variables. This follows  because dropping the time derivative term in the kinetic equation (\ref{e.lbtf}), the only difference to noninteracting case is that the distribution $\delta n_{\bm p}$ is replaced by $\delta\bar n_{\bm p}$. The boundary conditions (\ref{e.spbcu})-(\ref{e.dbcn3}) and the observables (\ref{e.n3dens})-(\ref{e.currt}) also are functions of $\delta\bar n_{\bm p}$ and they depend on the interaction parameters $m^*$, $\alpha$, $D$ and $F_l$'s in a simple scaling manner, if at all. 

The force applied to slowly moving objects in a Fermi liquid in the ballistic limit was calculated in Ref.\ \onlinecite{VT06}.
All the results presented in Ref.\ \onlinecite{VT06}  [Equations (13)-(19)] concern the time-indepent case. Therefore, their independence of the interactions parameters $m^*$, $\alpha$, $D$ and $F_l$ follows most simply from the general argument of the preceding paragraph.

In time dependent problems, the interactions have important effect  through the $\delta\bar \epsilon$-term in the kinetic equation (\ref{e.lbtf}). For small frequencies, however, the coupling to the superfluid motion can be neglected, as argued in Sec.\ \ref{s.eli}. Thus at low frequencies the response of Fermi-Bose liquid to external perturbation is the sum of independent superfluid and Fermi-liquid responses.

The theory formulated here is applied to calculate the force on a vibrating cylinder in Refs.\ \onlinecite{VTletter}, \onlinecite{VTres}, \onlinecite{VT06}, and \onlinecite{VT09}.


\end{document}